\documentclass[final,3p,times,authoryear]{elsarticle}
\usepackage{amssymb}
\usepackage{lipsum}
\usepackage{natbib}
\usepackage{amsmath}
\usepackage{threeparttable}
\usepackage{makecell}
\usepackage{xcolor}
\usepackage{float}

\usepackage{hyperref}
\hypersetup{
	    colorlinks=true, 
	    linktoc=all,     
	urlcolor  = blue,
	citecolor = blue,
	    linkcolor=blue,  
	}

\journal{Nuclear Physics A}

\begin{document}

\title{Shell-model study for $^{204-210}$Tl isotopes and core excitations across the $Z = 82$ and $N = 126$ shell gaps}

\author{Bharti Bhoy $^{1,2}$, Praveen C. Srivastava$^{1}$,\footnote{Corresponding author: praveen.srivastava@ph.iitr.ac.in}}
\address{$^{1}$ Department of Physics, Indian Institute of Technology Roorkee, Roorkee
247 667, India}
\address{$^{2}$Present affiliation: Department of Physics, University of Surrey, Guildford, GU2 7XH, UK.}

\date{\hfill \today}

\begin{abstract}

In the present work, the $^{204-210}$Tl isotopes have been investigated by performing large-scale shell-model calculations, including configurations allowing both neutron and proton core excitations across the $Z$ = 82 and $N$ = 126 shell gaps. Inspired by the recent high-spin experimental data, the structure of Tl isotopes has been studied for a considerably large model space.  The KHHE interaction has been used for $^{204-206}$Tl isotopes, KHH7B interaction for $^{204-210}$Tl isotopes, and additionally KHM3Y interaction has been used for $^{208}$Tl isotope. The core excitation has been performed using the KHH7B and KHM3Y interactions. The level spectra of $^{204-210}$Tl isotopes are comprehensively described and explained by multi-nucleon couplings of single-particle-hole orbitals within the valence space and by core excitations across $^{208}$Pb core. The well-known isomeric states are analyzed in terms of the shell model configurations. 

\end{abstract}

\maketitle

\section{Introduction}

Nuclei in the doubly magic $^{208}$Pb region, with the new incoming experimental data, provide one of the best testing grounds for the accuracy of the shell model calculations to reproduce and describe quantitatively experimentally established level structures. The progress in the experimental research has posed challenges for the verification of shell model calculations with the inclusion of more complex configurations that requires the participation of core excitations to form high-$j$ states. A significant contribution was found to revise the effective interactions in the experimental structure study of neutron-rich $^{206}$Hg, $^{204}$Pt (  $Z < $82 ) isotopes (e.g. \cite{fornal2,204Pt}) including proton holes. In the spherical shell model study of the $^{204}$Pt nucleus, a  modification of the two-body interaction for $N$=126, $Z <$ 82 was suggested. This modification improved the explanation of the levels and $\gamma$ transition rates influencing the evolution of nuclear structure toward the $r$-process waiting point nuclei. In this region, preliminary spectroscopy studies mainly concentrated on mercury isotopes and isomeric decay. In  $^{208}$Hg and $^{210}$Hg isotopes, with the yrast structures and isomeric states, a seniority-like structure was observed \cite{dahan,gottardo1}. In $^{207}$Hg \cite{Tang}, single-neutron states just above the magic number $N$ = 126 have been observed, which is very important for the shell-model. Similarly, $^{208}$Tl provides the proton-neutron two-body matrix elements for the shell-model calculations with one proton hole and one neutron outside $^{208}$Pb \cite{Nushellx2}. 

The deep-inelastic heavy-ion reactions technique used for discrete $\gamma$ spectroscopy is allowed to study the yrast structure at the neutron-rich side of the region in the vicinity of $^{208}$Pb \cite{rejmund,lane,fornal1,broda1,broda2}. Recently, remarkable progress was achieved in some of the well-studied nuclei, significantly extending the yrast structures to the high-$j$ value of the already observed excited states. In the less explored $N >$ 126, $Z <$ 82 region \cite{korten}, with the advancement of radioactive-beam facilities, $\gamma$-ray spectroscopy following internal decay of yrast excited states were observed for $^{208}$Tl \cite{208Rn} and $^{209}$Tl \cite{elle,bms,209Rn}, $\gamma$-ray transitions in $^{211,213}$Tl \cite{gottardo2} were also identified.

In this regard, a detailed shell model study of thallium isotopes could simplify the spectroscopic studies in the Pb region, herewith the presence of many high-spin isomers and states. It could shed some light and show whether the structure of the even higher-$j$ yrast states can be explained by the influence of core excitations. In the proximity of doubly magic nuclei Pb, the presence of many high-spin isomeric states is a beneficial feature for insight into spectroscopic studies. The occurrence of isomeric states provides comprehensive information on the precise ordering and accurate gap of adjacent single proton and neutron energy levels. In these nuclei near shell closure, the well-established isomeric states arise due to intrinsic specific proton or neutron excitation with a difference of large angular momentum. The appearance of isomers also manifests the energy favoured to generate the core-excited states with particular proton-neutron configurations. 

The structure of high-spin isomers near the $^{208}$Pb region is generally dominated by configurations involving high-$j$ nucleons occupying the $\nu i_{13/2}$ and $\pi h_{11/2}$ orbitals. Isomers at high excitation are more favourable in this region due to the presence of the opposite parity  $\pi h_{11/2}$ and $\nu i_{13/2}$ orbitals and high-spin states having multi-nucleon configurations. Therefore the isotopes of Tl ($Z$ = 81), lying just below and above the $N=126$ shell-closure in the Segré chart, are suitable candidates for observing the structure of high-spin states and isomers. The isotopes of Tl near the $N$ = 126 shell closure is possibly one of the least explored regions, with several unconfirmed high-spin and isomeric states \cite{204Tl,205Tl,Wilson,208Tl,bms,210Tl}. This region is interesting to study as the proton and neutron hole configurations get exhausted to generate the angular momentum, and the particle-hole excitations across the $Z$ = 82 and $N$ = 126 shell closure give rise to higher spin yrast states.

Two types of excitations are responsible for describing the structure of nuclei near the doubly closed-shell. One is the valence particle excitation, and the other is the coupling of single-particle valence nucleons to core excitations. 
A correlation between both excitations must be incorporated into the shell-model calculations to understand and describe the overall structure. For the high-energy core excited states across the $Z$ = 82 and $N$ = 126 closed shell, a larger model space with several orbitals below and above the shell gap is mandatory. Finding such a suitable effective interaction with manageable shell model calculations is very hard. In the shell-model calculations, the attractive proton-neutron residual interactions predominantly control the formation of the core-excited states \cite{Poletti,Byrne}. This attractive interaction enhances the chances of nucleon excitation out of the core to high-$j$ orbitals above the Fermi surface to achieve large angular momentum.

The present study aims to perform comprehensive shell-model calculations of $^{204-210}$Tl isotopes to discuss structural variation in terms of core excitations across $Z$ = 82 and $N$ = 126 closed shell. The energy spectrum and electromagnetic properties are calculated and compared with the available experimental data. There is no systematic shell-model study available for the Tl chain in the literature. In this work, we have included the competition between the valence single particle state excitations and core excitations for both protons and neutrons. The latter excitation has been guessed in many works. However, it has not been comprehensively demonstrated in terms of spherical shell-model calculations.

This paper is divided into the following sections. In Section \ref{section2}, we have discussed briefly the interaction used in our calculations. In Section \ref{section3}, we present our calculated results of energy spectra and electromagnetic properties for $^{204-210}$Tl isotopes. Finally, we summarize our results and conclude the paper in section \ref{summary}.

\section{Formalism: Shell Model Space and Interactions}
\label{section2}

The nuclear shell-model Hamiltonian can be expressed in terms of single-particle energies and two nucleon interactions,
	
	\begin{equation}
		H = \sum_{\alpha}\epsilon_{\alpha}c_\alpha^{\dagger}c_\alpha+\frac{1}{4}\sum_{\alpha \beta \gamma \delta JT}\langle j_\alpha j_\beta |V|j_\gamma j_{\delta} \rangle_{JT}c_\alpha^{\dagger}c_\beta^{\dagger}c_\delta c_\gamma
	\end{equation}
	
	where $\alpha=\{n,l,j,t\}$ stand for the single-particle orbitals and $\epsilon_{\alpha}$ are corresponding single-particle energies. $c_{\alpha}$ and $c_{\alpha}^{\dagger}$ are the fermion annihilation and creation operators. $\langle j_\alpha j_\beta |V|j_\gamma j_\delta \rangle_{JT}$ denote the antisymmetrized two-body matrix elements coupled to spin $J$ and isospin $T$.

Systematic studies have been carried out to understand the structure of Tl isotopes with $A$ = 204-210 within a more extensive model space containing the proton/neutron-core excitations across the closed $Z$ = 82 and $N$ = 126 shell.
The observed and calculated excited level energies of $^{204-210}$Tl isotopes have been investigated by considering core excitation and evaluating the inclusion of certain mixing in the framework of the shell-model.

To diagonalize the matrices, the OXBASH \cite{Oxbash}, NUSHELLX \cite{Nushellx1, Nushellx2} and KSHELL \cite{Kshell} codes have been employed for the shell-model calculations. Here we have taken three interactions, KHHE \cite{Warburton1}, KHH7B \cite{pbpop} and KHM3Y \cite{M3Y} for different sets of isotopes. Our focus is mainly on the application of KHH7B on the whole Tl chain and KHM3Y interaction for neutron number above $N=126$; as core excitation has been performed using these interactions, KHHE interaction has been used only for the comparison. The OXBASH and NUSHELLX shell-model codes are used to perform the pure core excitation calculations. For the calculations corresponding to without core excitation, the KSHELL code is used. 

For $^{204-206}$Tl isotopes, both KHHE and KHH7B interactions have been used, and for $^{207-210}$Tl isotopes, KHH7B interaction has been used.  The KHM3Y interaction has been used for $^{208}$Tl isotope. We have performed different sets of shell model calculations for different ranges of isotopes. The first set is the $t=0$ (no core excitations), which has been performed for all the isotopes. The second set is the $t=0+1$ (mixing between the $t=0$ and $t=1$), allowing both proton and neutron excitations across the $Z$ = 82 and $N$ = 126 shell, which is performed for the entire range of isotopes excluding $^{207}$Tl. And the third one is the $t=1$ ( $2h-1p$ core excitations), only for $^{207}$Tl. Here mixing between $t=0$ valence and core-excited configurations has been excluded, as there is only one valence proton hole in $^{207}$Tl, which does not need correlations, and experimental binding energies and single-particle/hole excitation energies are used in the interaction considered.

The KHHE \cite{Warburton1, Warburton2} interaction was constructed based on holes in a $^{208}$Pb core. The model space here consists of $0g_{7/2}$, $1d_{5/2}$, $1d_{3/2}$, $2s_{1/2}$, $0h_{11/2}$ proton orbitals and $0h_{9/2}$, $1f_{7/2}$, $1f_{5/2}$, $2p_{3/2}$, $2p_{1/2}$, $0i_{13/2}$ neutron orbitals.  The effective realistic residual interaction of Kuo and Herling \cite{Kuo1, Kuo2} was derived from a free nucleon-nucleon potential of Hamada and Johnston \cite{Hamada} with renormalization due to the finite extension of model space by the reaction matrix techniques developed by Kuo and Brown \cite{Kuo3}. In the present work, shell model calculations with KHHE interaction have been performed without truncation. The full-fledged calculation with KHHE interaction is sufficient to explain low-lying states in $^{204-206}$Tl isotopes. However, as discussed later, we need core excitation to explain high-lying states.

\begin{figure}
	\begin{center}
		\includegraphics[width=105mm,height=130mm,keepaspectratio]{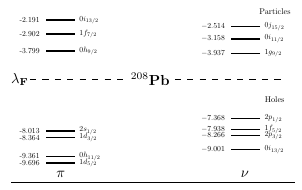}
	\end{center}
	\caption{The model space corresponding to KHH7B interaction.}
	\label{interaction}
\end{figure}

The KHH7B interaction consists of 14 orbitals having the proton orbitals between $Z$ = 58-114: $1d_{5/2}, 1d_{3/2}, 2s_{1/2}, 0h_{11/2}$, $0h_{9/2}, 1f_{7/2}, 0i_{13/2}$ and neutron orbitals between $N$ = 100-164:  $1f_{5/2}, 2p_{3/2},  2p_{1/2}, 0i_{13/2}$, $1g_{9/2}, 0i_{11/2}, 0j_{15/2}$. The KHH7B residual interaction used by Poppelier and Glaudemans \cite{pbpop} is the Surface Delta Interaction (SDI), which is the schematic interaction but gives the same results as the Kuo-Herling matrix elements \cite{Kuo1, Kuo2}.
In the KHH7B interaction, the cross-shell two-body matrix elements (TBMEs) were
generated by the G-matrix potential (H7B) \cite{Hosaka}. At the same time, the proton-neutron, hole-hole, and particle-particle TBMEs are taken from Kuo-Herling interaction \cite{Kuo1} as modified in Ref. \cite{Warburton1}. The orbitals used for KHH7B interaction are shown in Fig. \ref{interaction}.

The KHM3Y interaction consists of 24 orbitals with the proton orbitals $0g_{7/2}$, $1d_{5/2}$, $0h_{11/2}$, $1d_{3/2}$, $2s_{1/2}$ below $Z$=82 and $0h_{9/2}$, $1f_{7/2}$, $0i_{13/2}$, $1f_{5/2}$, $2p_{3/2}$, $2p_{1/2}$ above, and the neutron orbitals $0i_{13/2}$, $2p_{3/2}$, $1f_{5/2}$, $2p_{1/2}$, $0h_{9/2}$, $1f_{7/2}$ below $N=126$ and $1g_{9/2}$, $0i_{11/2}$, $0j_{15/2}$, $1g_{7/2}$, $2d_{5/2}$, $2d_{3/2}$, $3s_{1/2}$ above. Same as the KHH7B interaction, the proton-neutron interactions are based on the Kuo–Herling interaction \cite{Kuo1} as modified in Ref. \cite{Warburton1}, and the cross-shell two-body matrix elements (TBMEs) are based on the M3Y interaction \cite{M3Y}.

Previously, we have reported an extensive shell model calculation using KHH7B interaction for Rn isotopes \cite{Rn}. In this work, we found that core excitation is essential for describing high-lying states and mixing the single-particle states and core excitation above $Z=82$ and $N=126$ shell closure. Due to computational restriction, we could not achieve the core excited calculation for Rn isotopes. In continuation of our earlier work  \cite{Rn} and the availability of several experimental data, we have chosen Tl isotopes near doubly magic $^{208}$Pb to see the importance of core excitation across the closed shell.

In the case of $t=1$ and $t=0+1$ calculation from KHH7B interaction, the protons and neutrons are distributed to occupy the orbitals below $Z = 82$ and $N=126$ without any truncation for $^{204-210}$Tl. Above $Z = 82$ and $N=126$ for core excitations, one proton is allowed to excite in any one of the $0h_{9/2}, 1f_{7/2}, 0i_{13/2}$ orbitals, and one neutron is allowed to excite to any of the $1g_{9/2}, 0i_{11/2}, 0j_{15/2}$ orbitals (in addition to the extra valence neutrons above $^{207}$Tl). To make the shell-model calculations feasible and due to the huge dimension, for the calculation of $^{208}$Tl isotope from KHM3Y interaction, truncations have been applied for the orbitals below $N=126$. Only one neutron across core excitation is allowed (without proton excitation), and the lowest $h_{9/2}$, $f_{7/2}$ orbitals are closed to make a hypothetical $N$ = 100 core. In the case of $t=0$ calculation, for $^{204-207}$Tl, all the protons and neutrons are distributed without truncation in the orbitals below $Z = 82$ and $N=126$ shell closure. For $^{208-210}$Tl, proton orbitals are opened below shell closure. In contrast, neutron orbitals are completely filled to avoid excitation. For the three neutron orbitals above the shell closure, neutrons can occupy only up to the valence neutron(s) in each isotope as the mass number increases. Computationally it is challenging to perform shell model calculations without truncation in the Pb region.

\section{Results and Discussion} 
\label{section3}

This section discusses the calculated energy levels and electromagnetic transition properties and compares with the experimental data for $^{204-210}$Tl isotopes. The theoretical calculation has been performed up to at least those observed yrast energy levels from which core excitation has been observed. We have predicted the energy and spin value for those isotopes where core excited states have not been observed, which can be a candidate for the core excited state. 
The configurations for the core excited states are listed in Table \ref{t_Con_Tl} and their calculated energies are compared with the experimental data.

\subsection{Even Tl Isotopes}

\begin{figure}[h]
	\begin{center}
		\includegraphics[width=105mm,height=130mm,keepaspectratio]{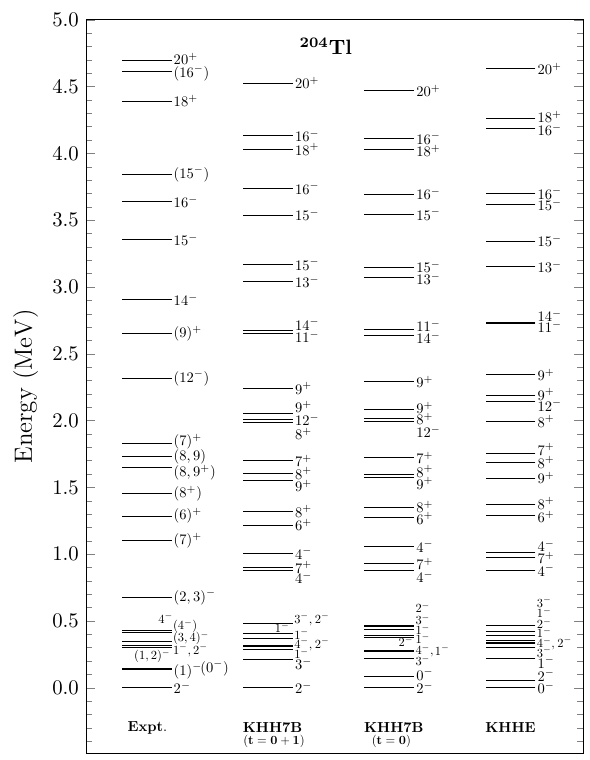}
	\end{center}
	\caption{ Comparison of calculated and experimental \cite{NNDC,204Tl}  excitation energy spectra for $^{204}$Tl.}
	\label{204Tl}
\end{figure}

\begin{figure}[h]
	\begin{center}
		\includegraphics[width=105mm,height=130mm,keepaspectratio]{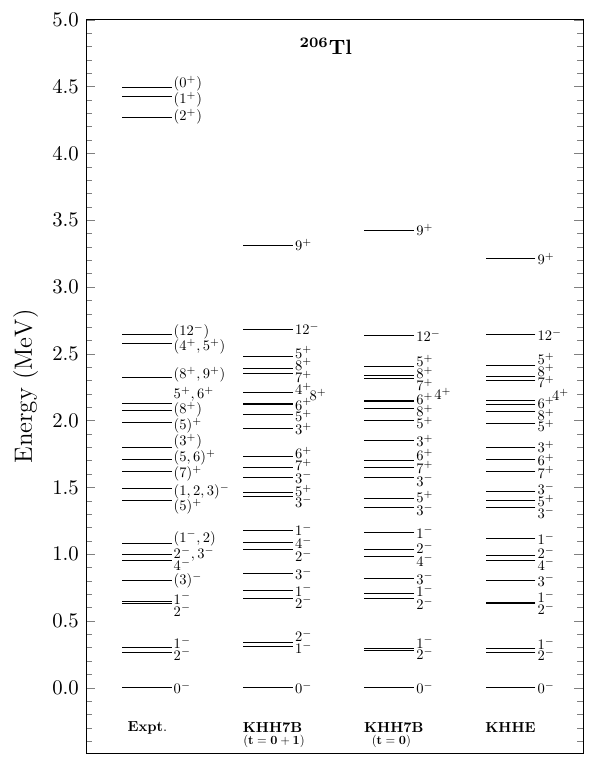}
	\end{center}
	\caption{ Comparison of calculated and experimental \cite{NNDC} excitation energy spectra for $^{206}$Tl.}
	\label{206Tl}
\end{figure}

\begin{figure}[h]
	\begin{center}
		\includegraphics[width=105mm,height=130mm,keepaspectratio]{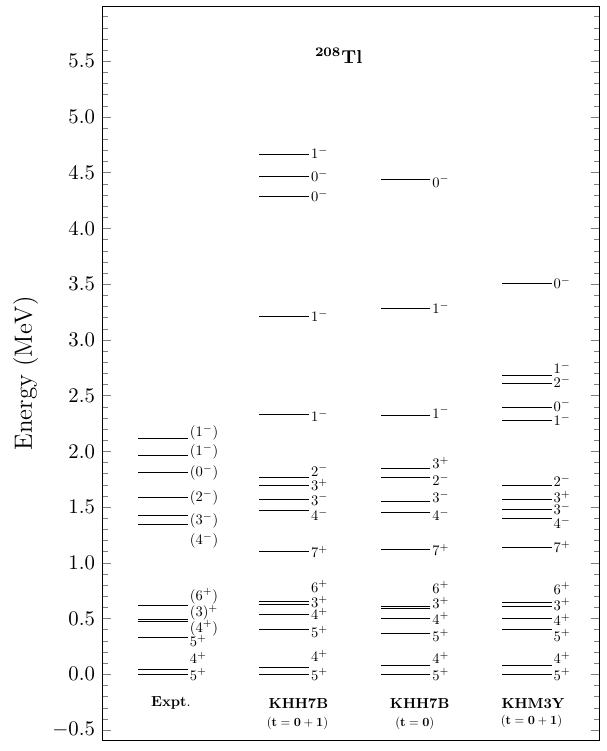}
	\end{center}
	\caption{ Comparison of calculated and experimental \cite{NNDC,208Tl} excitation energy spectra for $^{208}$Tl.}
	\label{208Tl}
\end{figure}

\begin{figure}[h]
	\begin{center}
		\includegraphics[width=105mm,height=130mm,keepaspectratio]{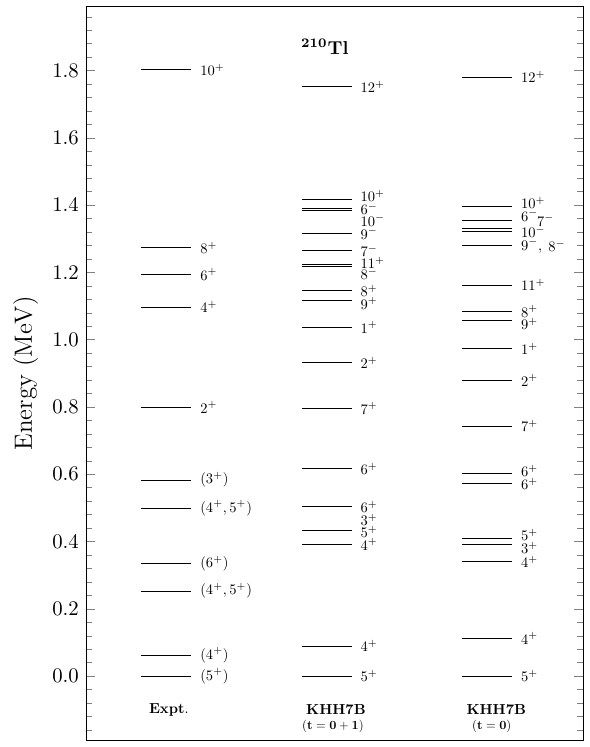}
	\end{center}
	\caption{ Comparison of calculated and experimental \cite{NNDC,210Tl} excitation energy spectra for $^{210}$Tl.}
	\label{210Tl}
\end{figure}


{\bf $^{204}$Tl:} Comparison of the experimental excitation energies for the positive and negative parity states of $^{204}$Tl with the predictions of shell model calculations are shown in  Fig. \ref{204Tl} up to the $20^+$ state. The high-spin structure of the $^{204}$Tl isotope was studied in Ref. \cite{204Tl} up to 11.2 MeV excitation energy and $J$ = 30 spin range. The energy levels of $^{204}$Tl established up to the highest spin $22^-$ state can be well understood by shell-model configurations involving four valence holes without core excitation. The ground state is mixed with the largest contribution 61.27\% of the $\pi s_{1/2}^{-1} \nu f_{5/2}^{-1}$ configuration. The well-established four yrast isomeric states $J^\pi$ = $7^+$, $12^-$, $18^+$, and  $22^-$ can be explained as arising from the coupling of one-proton and three-neutron holes. Our calculation confirms the predicted spin and parity of the four experimentally established isomeric states from both the interactions with the main configuration and the contribution discussed below with the KHH7B interaction. The structure of the lowest-lying, $7^+$ isomer at 0.902 MeV is a mixed state with a predominant $\pi s_{1/2}^{-1} \nu i_{13/2}^{-1}$ configuration with 67.27\% contribution. For the $12^-$ isomer at 2.008 MeV, the spin-parity is straightforward as arising from the $\pi h_{11/2}^{-1} \nu i_{13/2}^{-1}$ configuration with 80.54\% contribution, which is the highest spin state available when the two remaining neutron holes are paired to spin-zero. The structure of the intermediate $18^+$ isomer at 4.030 MeV arise with a $\pi h_{11/2}^{-1} \nu i_{13/2}^{-2} p_{1/2}^{-1}$ predominant configuration with 82.03\% probability.
The isomer $22^-$ at 5.978 MeV is purely arising from the $\pi h_{11/2}^{-1} \nu i_{13/2}^{-3}$ configuration with 83.78\% contribution (not shown in figure). 
For this highest-lying isomeric state, the spin reaches the maximum value for four valence holes. Above the $22^-$ isomeric state, core excitations are required for the structure of the yrast levels to induce any higher spin excitation. As shown in  Fig. \ref{204Tl}, the shell model calculations within both sets corresponding to KHH7B agree with the low-lying levels. The small differences between the two sets of calculations for the low-lying levels indicate that the $t=0$ is reasonable to interpret the low-lying level structures of $^{204}$Tl. However, as discussed above, the core excitations must be considered for the high-spin states. In Table \ref{t_Con_Tl}, the configuration of the core excited states is given up to $30^+_1$ state. The configurations of neutron-core excitations appear at the negative parity states from the $23^-$ at 10.2 MeV ( 7.052 MeV experimentally). At this energy, it is energetically more favourable for $^{204}$Tl to increase angular momentum by breaking the closed core than by exciting more protons and neutrons to the $\pi h_{9/2}$ and $\nu g_{9/2}$,$\nu i_{11/2}$, $\nu j_{15/2}$ orbitals. It should be noted that the negative parity states are dominated by neutron core excitation, except for the $26^-_1$ state. For the positive parity states, the configuration indicates strong competition between the pure proton core excitations ($\pi h_{11/2}^{-2} h_{9/2}^{1}$)  and neutron-core excitations.


{\bf $^{206}$Tl:} Comparison of the experimental excitation energies for the positive and negative parity states of $^{206}$Tl with the predictions of shell model calculations are shown in  Fig. \ref{206Tl}. The energy levels of $^{206}$Tl can be well understood by shell-model configurations involving two valence holes. For $^{206}$Tl, the energy spectrum with core excitation is not straightforward in terms of increasing spin. 
The ground state $0^-$ is coming with the 77.01\% contribution of the $\pi s_{1/2}^{-1} \nu p_{1/2}^{-1}$ configuration. The three yrast isomeric states $J^\pi$ = $5^+$, $7^+$, and $12^-$ can be explained as arising from the one-proton and one-neutron hole. The structure of the lowest-lying, $5^+$ isomer at 1.465 MeV arises with a predominant $\pi h_{11/2}^{-1} \nu p_{1/2}^{-1}$ configuration with 71.12\% contribution. The structure of the $7^+$ isomer at 1.646 MeV arises with a $\pi s_{1/2}^{-1} \nu i_{13/2}^{-1}$ predominant configuration with 75.74\% probability. For the $12^-$ isomer at 2.686 MeV, the spin-parity is straightforward as arising from the $\pi h_{11/2}^{-1} \nu i_{13/2}^{-1}$ configuration with 83.47\% contribution, which is the highest spin state available when the two remaining neutron holes are paired. Above the $12^-$ isomeric state, core excitations are required for the structure of the yrast levels to induce any higher spin/energy excitation except for the calculated $9^+$ state, which is not a core excited state and is lying higher than the $12^-$ state ($\approx$ 1 MeV higher than the experimental data). One can see from the calculation for $^{204-206}$Tl isotopes that for $N < $ 126, the KHH7B interaction is giving the same results as the KHHE interaction. So, to calculate the nuclei above and below  $Z =$ 82, $N =$ 126 region, the KHH7B interaction is better and sufficient than the limited model-spaced interaction like KHHE. The KHH7B interaction is also more suitable because to explain the structure of the high spin state around $^{208}$Pb, the core excitation and mixing of high-lying orbitals to lower single-particle states are essential. Similar to $^{204}$Tl, the low-lying levels of $^{206}$Tl can be reasonably interpreted with the $t=0$. However, the core excitations must be considered, as discussed above for the higher energy states above the $12^-$ isomeric state. In Table \ref{t_Con_Tl}, the configurations of the core excited states are given. The configurations of the lowest neutron-core excitations appear for the $3_2^+$ at 5.885 MeV (2.868 MeV experimentally). 
 The lowering of energy for the core polarization might be due to the one neutron hole in the $N$ = 126 shell by which the core softens. It is reflected from the configuration as all the core excited states arise from the neutron excitation only.


{\bf $^{208}$Tl:} Comparison of the experimental excitation energies for the positive and negative parity states of $^{208}$Tl with the predictions of shell model calculations are shown in  Fig. \ref{208Tl}. Two interactions KHH7B and KHM3Y, have been used to calculate energy spectra of $^{208}$Tl. There is a good agreement between the calculated and experimental energy spectra. The energy levels of $^{208}$Tl can be well understood by shell-model configurations involving one-proton-hole and one-neutron-particle without core excitation in the valence space only. For some of the energy levels, we need $2p2h$ core excitation from the low-lying orbitals across the $N$=126 shell closure. Most of the calculated low-lying yrast and near-yrast states from the KHH7B interaction are constructed with leading $ph$ configuration  $\pi (s,d) \nu g_{9/2}$ and $ \pi h_{11/2} \nu j_{15/2}$ for positive parity states and $ \pi h_{11/2} \nu (g_{9/2}, i_{11/2})$ for negative parity states. While from the KHM3Y interaction, the leading $ph$ configurations are also based on these sets of orbitals but with different structures, especially for the states considered $\nu j_{15/2}$ orbital is not playing any role, which is important in KHH7B interaction. For $^{208}$Tl, when we extend our model space with KHM3Y interaction, it shows an increase in the role of low-$j$ orbitals. The ground state arises with the $\pi s_{1/2}^{-1} \nu g_{9/2}^{1}$ configuration from both the interactions. Up to 1 MeV excitation energy, the states are generated with the same configuration from both interactions. After this energy, KHH7B interaction generates states only with high-$j$ proton and neutron orbitals, which is not the case in KHM3Y interaction. In $^{208}$Tl, the only isomeric state $0^-$ at 1.807 MeV is of core excited nature \cite{208Tl}. In our calculation, the KHH7B interaction could not correctly reproduce the energy of the newly identified levels as reported in Ref. \cite{208Tl}. Also, the wave function of the $0^-$ state is $\pi h_{11/2}^{-1} \nu i_{11/2}^{1}$ and the calculated energy is 4.290 MeV. From the KHM3Y interaction the configuration of the $0^-$ state is reproduced as $\pi s_{1/2}^{-1} \nu g_{9/2}^2 p_{1/2}^{-1}$(90.99\%) with the calculated energy 2.392 MeV. In $^{208}$Tl, a relatively low excitation energy can be seen, which can be explained by the strong pairing of $\nu g_{9/2}^2$, which is favourable to break the closed shell $N$=126. In KHH7B interaction, the core-excited state starts at 4.668 MeV and the $\pi s_{1/2}^{-1} \nu g_{9/2}^2 p_{1/2}^{-1}$ configuration is for the $0_2^-$ and $1_3^-$ states, whereas in KHM3Y it is the configuration for $0_1^-$ and $1_2^-$ states. This comparison of two interactions shows the importance of including low-$j$ orbitals and extension of model space beyond $\nu j_{15/2}$ orbital and $N = 126$ shell closure. Here we can see that to reproduce the energy levels and wave-functions correctly, we should consider the pairing and interaction among the orbitals between $N$ = 126-184.


{\bf $^{210}$Tl:} Comparison of the experimental excitation energies for the positive and negative parity states of $^{210}$Tl with the predictions of shell model calculations are shown in  Fig. \ref{210Tl}. In $^{210}$Tl, all the low-lying states up to 0.7 MeV are tentative, including the ground state. The calculation predicts ground state as  $J^\pi$ = $5^+$, although experimentally it is tentative. All of the calculated positive parity states, up to a 1.5 MeV excitation energy, arise predominantly from the $\pi s_{1/2}^{-1} \nu g_{9/2}^3$ configuration in both sets of calculation. This configuration allows spin couplings up to the $J$ = 11. The $6^+_1$ and $6^+_2$ states are not in this group and arise from different configurations. The $6^+_1$ state is due to $\pi s_{1/2}^{-1} \nu i_{11/2}^{1}g_{9/2}^2 $ configuration and $6^+_2$ state is due to $\pi d_{3/2}^{-1} \nu g_{9/2}^3$, from the core-excited calculation. This configuration has been flipped for both states in the calculation without core excitations. All the negative parity states in both sets of calculation arise predominantly from the $\pi h_{11/2}^{-1} \nu g_{9/2}^3$ configuration. The $8^-$ state arise predominantly from the $\pi s_{1/2}^{-1} \nu g_{9/2}^2 j_{15/2}^{1}$ configuration. In the calculated state, no core excitations can be seen. 

\subsection{Odd Tl Isotopes}

\begin{figure}[h]
	\begin{center}
		\includegraphics[width=105mm,height=130mm,keepaspectratio]{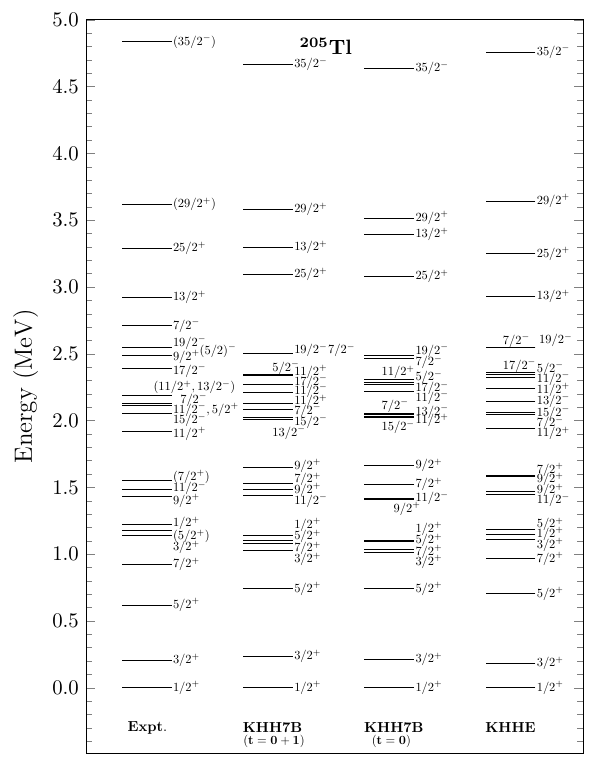}
	\end{center}
	\caption{ Comparison of calculated and experimental \cite{NNDC,205Tl} excitation energy spectra for $^{205}$Tl.}
	\label{205Tl}
\end{figure}

\begin{figure}[h]
	\begin{center}
		\includegraphics[width=105mm,height=130mm,keepaspectratio]{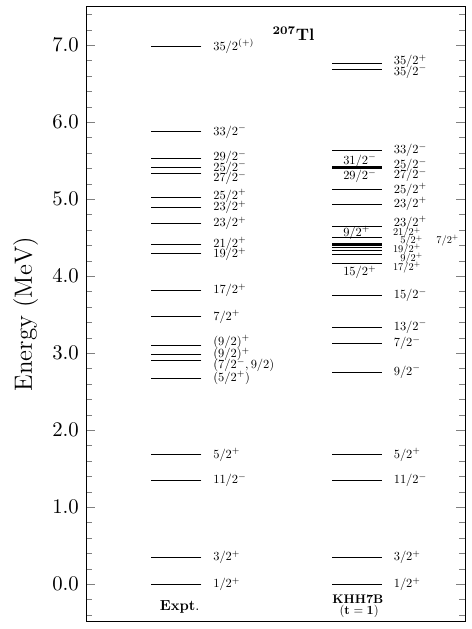}
	\end{center}
	\caption{ Comparison of calculated and experimental \cite{NNDC,Wilson} excitation energy spectra for $^{207}$Tl.}
	\label{207Tl}
\end{figure}

\begin{figure}[h]
	\begin{center}
		\includegraphics[width=105mm,height=130mm,keepaspectratio]{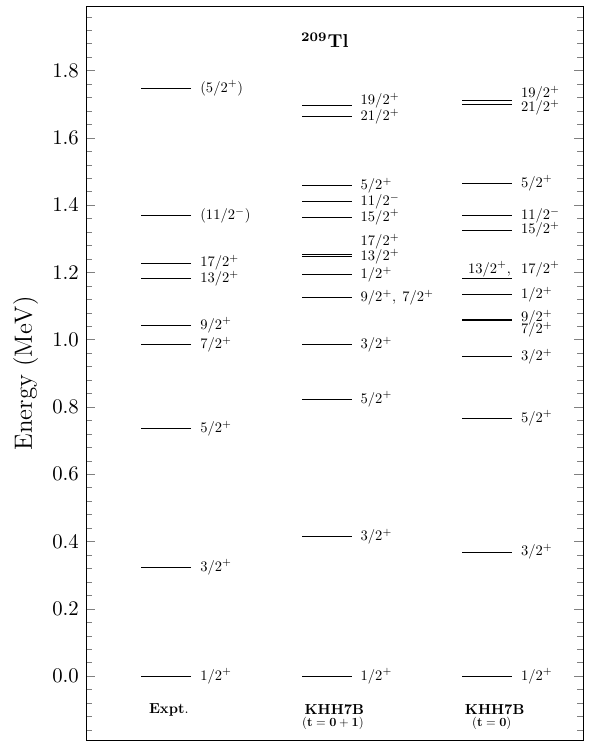}
	\end{center}
	\caption{ Comparison of calculated and experimental \cite{NNDC,bms} excitation energy spectra for $^{209}$Tl.}
	\label{209Tl}
\end{figure}


{\bf $^{205}$Tl:} Comparison of the experimental excitation energies for the positive and negative parity states of $^{205}$Tl with the predictions of shell model calculations are shown in  Fig. \ref{205Tl} up to $35/2^-$ state, without core excitation involving three valence holes. New high-spin states of the $^{205}$Tl isotope were identified in Ref. \cite{205Tl} and above the known isomer  $25/2^+$ with 2.6 $\mu$s half-life.
As shown in Fig. \ref{205Tl}, the shell model calculations within the two sets are in good agreement for the low-lying states. Same as $^{204}$Tl, the small differences between the two sets of calculations for the low-lying levels indicate that the $t=0$ is reasonable to interpret the structures of $^{205}$Tl. However, for the high-spin states, the core excitations must be considered. The energy levels of $^{205}$Tl established above $29/2^+$ for the positive parity state and $35/2^-$ for the negative parity state can be well understood by shell-model configurations involving three valence holes with the core excitation. The ground state arises from the $\pi s_{1/2}^{-1}  $ configuration with the largest contribution of 72.22\%. The well-established two yrast isomeric states $J^\pi$ = $25/2^+$ and $35/2^-$  can be explained as arising from the one-proton and two-neutron holes couplings to high-spin states. The structure of the $25/2^+$ isomer is with a predominant $\pi h_{11/2}^{-1} \nu p_{1/2}^{-1} i_{13/2}^{-1}$ configuration with 80.8\% contribution. The continuation of the yrast line above this three-hole isomeric state is obtained when the $\nu p_{1/2}$ neutron hole is replaced by the next accessible $\nu f_{5/2}$ and $\nu i_{13/2}$ holes. For the $35/2^-$ isomer at 4.667 MeV (calculated energy), the spin-parity is straightforward as arising from the $\pi h_{11/2}^{-1} \nu i_{13/2}^{-2}$ configuration with 83.68\% contribution. Our calculation confirms the $29/2^+$ state with $\pi h_{11/2}^{-1} \nu f_{5/2}^{-1} i_{13/2}^{-1}$ configuration, which is the highest positive spin state available when the three neutron holes are paired due to maximum available high-spin orbital. Above the $29/2^+$ state, core excitations are required for the structure of the positive parity yrast levels to induce any higher spin excitation. Although $37/2^-$ state can be formed without core excitation, it is arising from $\pi h_{11/2}^{-1} \nu f_{5/2}^{-1} p_{1/2}^{-1} i_{13/2}^{-1} g_{9/2}^{1}$ configuration with one neutron excited outside the core to $\nu g_{9/2}$ orbital. This may be due to the energy required to construct a $37/2^-$ state might be sufficient to break the neutron core. Above the $35/2^-$ isomer, the $41/2^+$ yrast state (experimentally at 7.092 MeV) was interpreted as arising from the coupling of valence three holes $\pi h_{11/2}^{-1} \nu i_{13/2}^{-2}$  of the $35/2^-$ isomer to the octupole vibration ( $3^-$ core excitation) of the $^{208}$Pb core \cite{205Tl}. The configuration of this state (calculated at 10.246 MeV energy) is $\pi h_{11/2}^{-1} \nu p_{1/2}^{-1} i_{13/2}^{-2} g_{9/2}^{1}$ with 72.22\% contribution. In Table \ref{t_Con_Tl}, the core excited states configuration is given up to the $41/2$ state for both parities. All the states considered in the $^{205}$Tl are arising due to neutron-core excitations except for the $41/2^-$ state. The energy where the core excitations begin to occur is 8.915 MeV, favourable for $^{205}$Tl to increase angular momentum by breaking the closed core accomplished by exciting more proton and neutron to the $\pi h_{9/2}$ and $\nu g_{9/2}$,$\nu i_{11/2}$, $\nu j_{15/2}$ orbitals.

{\bf $^{207}$Tl:}  Comparison of the experimental excitation energies for the positive and negative parity states of $^{207}$Tl with the predictions of shell model calculations are shown in  Fig. \ref{207Tl} up to $35/2$ spin. New high-spin yrast and near-yrast states of the $^{207}$Tl isotope were identified in Ref. \cite{Wilson}, above the known isomer with 1.33 s for $11/2^-$, and the level scheme was established up to $35/2^{(+)}$ spin at 6.984 MeV. The low-energy states of $^{207}$Tl are characterised by single proton-hole structures, with one proton-hole below the $Z$ = 82 magic number and closed $N$ = 126 neutron-shell. In contrast, the higher energy states arise due to the breaking of the core with a simple structure. The energy levels of $^{207}$Tl can be well understood by shell-model configurations involving two valence holes and one neutron breaking the core. The four lowest energy states interpreted by the single-proton holes are the following. The ground state has a $\pi s_{1/2}^{-1}$ configuration with 100$\%$ contribution, and next excited states after this are $\pi d_{3/2}^{-1}$(100$\%$), $\pi h_{11/2}^{-1}$(100$\%$), and $\pi d_{5/2}^{-1}$(100$\%$). Other excited states above these states are expected to emerge from the coupling between the single particle states and the $3^-$(octupole state), $4^-$, and $5^-$ excitations observed in $^{208}$Pb. As shown in Fig. \ref{207Tl}, there is a reasonable agreement between the shell model calculation and the experimentally observed core excited states. The shell-model calculations correctly predict the ordering of these states while their energies are slightly compressed at higher spins.  The shell model calculation overestimated these states, which shows a different character than the core excited state, associated with the nature of collective octupole states. The structure of the $35/2^-$ state is with a predominant $\pi h_{11/2}^{-1} \nu i_{13/2}^{-1} i_{11/2}^{1}$ configuration with 94.44\% probability, which is the highest negative spin state available when the single proton and single neutron holes are paired to one neutron in the $i_{11/2}$ orbital. The continuation of the yrast states above this negative parity state can be achieved when more particle-hole excitations are considered, as $1p1h$ excitation has exhausted at the $35/2^-$ state with three particles. In Table \ref{t_Con_Tl}, the configuration of the core excited states is given up to the $35/2^+$. One can see from Table \ref{t_Con_Tl}, most of the core excited states are constructed on the high spin proton $ h_{11/2}$ orbital, except for the states  $J^\pi$ = $7/2^-$,  $9/2^-$, $13/2^-$, $15/2^-$, $15/2^+$, $5/2^+$, $21/2^+$, and $23/2^+$.

\begin{table}[H]
	\begin{center}
		\caption{Configurations and excitation energies with KHH7B interaction for core-excited states in Tl isotopes with the probability of the dominant component of the configuration in comparison with the experimental excitation energies \cite{NNDC,204Tl,205Tl,Wilson}.}
		\label{t_Con_Tl}
		\resizebox{13.6cm}{!}{ 
			\begin{tabular}{cccccc}
				\hline
				\hline
				&  &  &   & &   \\
				
				Nucleus & $J^{\pi}$ & $E_x$ (Theo)  & $E_x$ (Exp) & Wave-function & Probability ($\%$) \\
				
				&  &  &   & &   \\
				\hline

				&  &  &   & &   \\
				$^{204}$Tl  & $23_1^-$ & 10.200 & 7.052 & $ \pi h_{11/2}^{-1} \nu f_{5/2}^{-1}p_{1/2}^{-1}i_{13/2}^{-2}g_{9/2}^{1}$ & 71.98     \\
				& $22_1^+$ & 10.352 &  & $\pi d_{3/2}^{-1} h_{11/2}^{-1} h_{9/2}^{1} \nu f_{5/2}^{-2}i_{13/2}^{-1}$ & 88.18     \\
				& $24_1^-$ & 10.377 &  & $ \pi h_{11/2}^{-1} \nu f_{5/2}^{-1}p_{1/2}^{-1}i_{13/2}^{-2}g_{9/2}^{1}$  & 77.31     \\
				& $25_1^-$ & 10.573 &  & $ \pi h_{11/2}^{-1} \nu f_{5/2}^{-1}p_{1/2}^{-1}i_{13/2}^{-2}g_{9/2}^{1}$ & 78.03     \\
				& $23_1^+$ & 10.818 &  & $ \pi h_{11/2}^{-1} \nu  i_{13/2}^{-2}j_{15/2}^{1}$ & 76.79     \\
				& $24_1^+$ & 10.953 &  & $ \pi h_{11/2}^{-1} \nu  i_{13/2}^{-2}j_{15/2}^{1}$ & 81.27     \\
				& $25_1^+$ & 11.195 & 8.447 & $ \pi h_{11/2}^{-1} \nu  i_{13/2}^{-2}j_{15/2}^{1}$ & 80.94     \\
				& $26_1^-$ & 11.416 &  & $\pi d_{3/2}^{-1} h_{11/2}^{-1} h_{9/2}^{1} \nu f_{5/2}^{-1}i_{13/2}^{-2}$ & 83.63     \\
				& $26_1^+$ & 11.466 &  & $\pi  h_{11/2}^{-2} h_{9/2}^{1} \nu p_{1/2}^{-1}i_{13/2}^{-2}$ & 65.21     \\
				& $27_1^+$ & 11.667 & 9.167 & $\pi  h_{11/2}^{-2} h_{9/2}^{1} \nu f_{5/2}^{-1}i_{13/2}^{-2}$ & 59.25     \\
				& $27_2^+$ & 11.795 & 9.430 & $\pi  h_{11/2}^{-1} \nu f_{5/2}^{-1} p_{1/2}^{-1} i_{13/2}^{-2} j_{15/2}^{1}$ & 56.84     \\
				& $28_1^+$ & 11.969 & 9.648  & $\pi  h_{11/2}^{-2} h_{9/2}^{1} \nu f_{5/2}^{-1}i_{13/2}^{-2}$ & 78.41     \\
				& $28_2^+$ & 12.072 & 10.421 & $\pi  h_{11/2}^{-1} \nu f_{5/2}^{-1} p_{1/2}^{-1} i_{13/2}^{-2} j_{15/2}^{1}$ & 83.41     \\
				& $29_1^+$ & 12.380 & 10.905 &  $\pi  h_{11/2}^{-2} h_{9/2}^{1} \nu f_{5/2}^{-1}i_{13/2}^{-2}$ & 66.88     \\
				& $30_1^+$ & 13.458 & 11.157 &  $\pi  h_{11/2}^{-1} \nu f_{5/2}^{-1}i_{13/2}^{-3} i_{11/2}^{1} $ & 91.54     \\
				\hline

				&  &  &   & &   \\
				$^{205}$Tl  & $31/2_1^+$ & 8.915 & & $\pi d_{3/2}^{-1} \nu f_{5/2}^{-1}p_{1/2}^{-1}i_{13/2}^{-1} g_{9/2}^{1}$ & 48.03     \\
				& $37/2_1^-$ & 9.372 & & $\pi h_{11/2}^{-1} \nu f_{5/2}^{-1}p_{1/2}^{-1}i_{13/2}^{-1} g_{9/2}^{1}$  & 80.44     \\
				& $33/2_1^+$ & 9.497 & & $\pi d_{3/2}^{-1} \nu f_{5/2}^{-1}p_{1/2}^{-1}i_{13/2}^{-1} i_{11/2}^{1}$  & 85.95     \\
				& $39/2_1^-$ & 9.626 & & $\pi h_{11/2}^{-1} \nu f_{5/2}^{-1}p_{1/2}^{-1}i_{13/2}^{-1} g_{9/2}^{1}$  & 68.24     \\
				& $35/2_1^+$ & 9.679 & & $\pi h_{11/2}^{-1} \nu  i_{13/2}^{-1} j_{15/2}^{1}$  & 69.59    \\
				& $37/2_1^+$ & 9.777 & & $\pi h_{11/2}^{-1} \nu  i_{13/2}^{-1} j_{15/2}^{1}$  & 77.55    \\
				& $39/2_1^+$ & 10.029 & & $\pi h_{11/2}^{-1} \nu  i_{13/2}^{-1} j_{15/2}^{1}$  & 85.98    \\
				& $41/2_1^-$ & 10.224 & & $\pi d_{3/2}^{-1} h_{11/2}^{-1} h_{9/2}^{1} \nu f_{5/2}^{-1}i_{13/2}^{-1} $  & 86.4    \\
				& $41/2_1^+$ & 10.246 & 7.091 & $\pi h_{11/2}^{-1} \nu p_{1/2}^{-1}i_{13/2}^{-2} g_{9/2}^{1}$  & 72.22    \\
				
				\hline

				&  &  &   & &   \\
				$^{206}$Tl  & $3_2^+$ & 5.885 & 2.868 & $\pi d_{3/2}^{-1} \nu   g_{9/2}^{1}$ & 41.49     \\
				& $2_1^+$ & 6.292 & 4.273 & $\pi s_{1/2}^{-1} \nu p_{3/2}^{-1} p_{1/2}^{-1} g_{9/2}^{1}$ & 37.41     \\
				& $7_2^-$ & 6.669 &  & $\pi s_{1/2}^{-1} \nu   j_{15/2}^{1}$ & 68.3    \\
				& $8_2^-$ & 6.674 & & $\pi s_{1/2}^{-1} \nu   j_{15/2}^{1}$ & 73.34     \\
				& $1_1^+$ & 6.709 & 4.426 & $\pi s_{1/2}^{-1} \nu p_{1/2}^{-1} i_{13/2}^{-1} j_{15/2}^{1}$ & 34.18  \\
				& $9_2^-$ & 6.848 &  & $\pi h_{11/2}^{-1} \nu   g_{9/2}^{1}$ & 76.22  \\
				& $6_2^-$ & 6.888 &  & $\pi h_{11/2}^{-1} \nu   g_{9/2}^{1}$ & 75.62 \\
				& $9_2^+$ & 6.955 &  & $\pi d_{3/2}^{-1} \nu f_{5/2}^{-1} p_{1/2}^{-1} i_{11/2}^{1}$ &  36.13 \\
				& $5_3^-$ & 6.959 &  & $\pi h_{11/2}^{-1} \nu   g_{9/2}^{1}$ & 76.36  \\
				& $10_2^-$ & 6.986 &  & $\pi h_{11/2}^{-1} \nu   g_{9/2}^{1}$ &  75.63 \\
				& $0_1^+$ & 7.014 & 4.491 & $\pi s_{1/2}^{-1} \nu p_{1/2}^{-1} i_{13/2}^{-1} j_{15/2}^{1}$ &  52.04 \\
				\hline

		\end{tabular}}
	\end{center}
\end{table}

\addtocounter{table}{-1}

\begin{table}[H]
	\begin{center}
		\leavevmode
		\caption{{Continuation.}}
		\resizebox{14.0cm}{!}{ 
			\begin{tabular}{cccccc}
				
				\hline
				&  &  &   & &   \\
				
				Nucleus & $J^{\pi}$ & $E_x$ (Theo)  & $E_x$ (Exp) & Wave-function & Probability ($\%$) \\
				
				&  &  &   & &   \\
				\hline

				& $11_2^-$ & 7.406 &  & $\pi s_{1/2}^{-1} \nu p_{1/2}^{-1} i_{13/2}^{-1} g_{9/2}^{1}$ & 68.3  \\
				& $12_2^-$ & 7.583 &  & $\pi s_{1/2}^{-1} \nu p_{1/2}^{-1} i_{13/2}^{-1} g_{9/2}^{1}$ & 68.67  \\
				& $10_1^+$ & 7.696 &  & $\pi d_{3/2}^{-1} \nu f_{5/2}^{-1} p_{1/2}^{-1} i_{11/2}^{1}$ & 79.90  \\
				& $11_1^+$ & 8.130 &  & $\pi h_{11/2}^{-1} \nu   j_{15/2}^{1}$ & 77.92  \\
				& $12_1^+$ & 8.130 &  & $\pi h_{11/2}^{-1} \nu   j_{15/2}^{1}$ & 76.04 \\
				
				\hline
				
				&  &  &   & &   \\
				$^{207}$Tl  & $9/2_1^-$ & 2.756 & 2.912 & $\pi h_{11/2}^{-1}$ & 81.15     \\
				& $7/2_1^-$ & 3.130 & 2.912 & $\pi s_{1/2}^{-1} \nu p_{3/2}^{-1} g_{9/2}^{1}$ & 34.9     \\
				& $13/2_1^-$ & 3.340 & & $\pi d_{3/2}^{-1} \nu p_{1/2}^{-1} g_{9/2}^{1}$ & 55.84     \\
				& $15/2_1^-$ & 3.750 & & $\pi s_{1/2}^{-1} \nu f_{5/2}^{-1} g_{9/2}^{1}$ & 91.56     \\
				& $15/2_1^+$ & 4.161 & & $\pi s_{1/2}^{-1} \nu p_{1/2}^{-1} j_{15/2}^{1}$ & 78.86     \\
				& $17/2_1^+$ & 4.278 & 3.813 & $\pi h_{11/2}^{-1} \nu p_{1/2}^{-1} g_{9/2}^{1}$ & 64.96     \\
				& $9/2_1^+$ & 4.335 & 2.985 & $\pi h_{11/2}^{-1} \nu p_{1/2}^{-1} g_{9/2}^{1}$ & 73.27     \\
				& $19/2_1^+$ & 4.379 & 4.293 & $\pi h_{11/2}^{-1} \nu p_{1/2}^{-1} g_{9/2}^{1}$ & 63.45     \\
				& $5/2_1^+$ & 4.395 & 2.912 & $\pi s_{1/2}^{-1} \nu i_{13/2}^{-1} g_{9/2}^{1}$ & 81.48     \\
				& $7/2_1^+$ & 4.407 & 3.474 & $\pi h_{11/2}^{-1} \nu p_{1/2}^{-1} g_{9/2}^{1}$ & 66.20     \\
				& $21/2_1^+$ & 4.431 & 4.418 & $\pi s_{1/2}^{-1} \nu i_{13/2}^{-1} g_{9/2}^{1}$ & 79.76     \\
				& $9/2_1^+$ & 4.498 & 3.104 & $\pi s_{1/2}^{-1} \nu i_{13/2}^{-1} g_{9/2}^{1}$ & 73.27     \\
				& $23/2_1^+$ & 4.645 & 4.683 & $\pi s_{1/2}^{-1} \nu i_{13/2}^{-1} g_{9/2}^{1}$ & 86.7     \\
				& $23/2_2^+$ & 4.937 & 4.896 & $\pi d_{3/2}^{-1} \nu i_{13/2}^{-1} g_{9/2}^{1}$ & 47.62     \\
				& $25/2_1^+$ & 5.122 & 5.026 & $\pi h_{11/2}^{-1} \nu f_{5/2}^{-1} g_{9/2}^{1}$ & 80.06     \\
				& $29/2_1^-$ & 5.403 & 5.524 & $\pi h_{11/2}^{-1} \nu i_{13/2}^{-1} g_{9/2}^{1}$ & 93.91     \\
				& $27/2_1^-$ & 5.409 & 5.328 & $\pi h_{11/2}^{-1} \nu i_{13/2}^{-1} g_{9/2}^{1}$ & 93.20     \\
				& $25/2_1^-$ & 5.416 & 5.409 & $\pi h_{11/2}^{-1} \nu p_{1/2}^{-1} j_{15/2}^{1}$ & 93.96     \\
				& $31/2_1^-$ & 5.427 & & $\pi h_{11/2}^{-1} \nu i_{13/2}^{-1} g_{9/2}^{1}$ & 95.41     \\
				& $33/2_1^-$ & 5.636 & 5.876 & $\pi h_{11/2}^{-1} \nu i_{13/2}^{-1} g_{9/2}^{1}$ & 95.49     \\
				& $35/2_1^-$ & 6.686 & & $\pi h_{11/2}^{-1} \nu i_{13/2}^{-1} i_{11/2}^{1}$ & 94.44     \\
				& $35/2_1^+$ & 6.769 & 6.985 & $\pi h_{11/2}^{-1} \nu i_{13/2}^{-1} j_{15/2}^{1}$ & 93.74     \\
				
				\hline 
				
				&  &  &   &    \\
				$^{208}$Tl  & $11_2^-$ & 6.190 & & $\pi d_{3/2}^{-1} \nu p_{1/2}^{-1} i_{11/2}^{1} g_{9/2}^{1}$ & 77.61     \\
				& $12_2^+$ & 6.496 & & $\pi s_{1/2}^{-2} h_{9/2}^{1} \nu g_{9/2}^{1}$ & 83.97     \\
				& $11_2^+$ & 6.514 & & $\pi s_{1/2}^{-2} i_{13/2}^{1} \nu g_{9/2}^{1}$ & 79.67     \\
				& $12_1^-$ & 6.576 & & $\pi d_{3/2}^{-1} \nu p_{1/2}^{-1} i_{11/2}^{1} g_{9/2}^{1}$ & 46.3     \\
				& $10_2^+$ & 6.819 & & $\pi s_{1/2}^{-2} h_{9/2}^{1} \nu j_{15/2}^{1}$ & 77.43     \\
				& $1_1^+$ & 6.864 & 13.600 & $\pi s_{1/2}^{-1} \nu p_{1/2}^{-1} i_{11/2}^{1} j_{15/2}^{1}$ & 79.76     \\
				& $0_1^+$ & 7.268 & 7.000 & $\pi d_{3/2}^{-1} \nu p_{1/2}^{-1} i_{11/2}^{1} j_{15/2}^{1}$ & 68.72     \\

				\hline
				\hline

		\end{tabular}}
	\end{center}
\end{table}



\begin{table}[H]
	\begin{center}
		\caption{The calculated (SM) electromagnetic transition strengths in W.u. for Tl isotopes from KHH7B interaction compared to the experimental data (Expt.) \cite{NNDC,204Tl,205Tl,Wilson,208Tl,bms,210Tl}
			using the effective charges $e_\pi = 1.5e$ and $e_\nu = 0.5e$.}
		\label{t_trans}
		\resizebox{9.55cm}{!}{ 
			\begin{tabular}{ccccc}
				\hline
				\hline
				
				&  &  &   &    \\
				Nucleus & Transition & $E/M (\lambda)$ &   Theo. ($t=0$)  &  Expt.    \\
				
				&  &  &   &   \\
				\hline

				&  &  &   &   \\
				$^{204}$Tl & $0_1^- \rightarrow  1_1^-$ & $M1$ & 0.485  &     \\
				& $0_1^- \rightarrow  2_1^-$ & $E2$ & 0.342  & 0.1     \\
							
				&  &  &   &    \\
				$^{205}$Tl & $3/2_1^+ \rightarrow  1/2_1^+$ & $M1$ & 0.00022 &  0.00051(10)        \\
				& $3/2_1^+ \rightarrow  1/2_1^+$ & $E2$ & 3.742 &  6.1(9)        \\
				& $5/2_1^+ \rightarrow  3/2_1^+$ & $M1$ & 0.489 &  0.27(5)        \\
				& $5/2_1^+ \rightarrow  3/2_1^+$ & $E2$ & 0.713 &  2.7(10)        \\
				& $5/2_1^+ \rightarrow  1/2_1^+$ & $E2$ & 2.626 &  6.0(12)        \\
				& $7/2_1^+ \rightarrow  3/2_1^+$ & $E2$ & 2.59 &         \\
								& $15/2_1^- \rightarrow  11/2_1^-$ & $E2$ & 0.865 &          \\
			
				&  &  &   &    \\
				$^{206}$Tl & $2_1^- \rightarrow  0_1^-$ & $E2$ & 1.192 &  2.22(14)      \\
				& $1_1^- \rightarrow  0_1^-$ & $M1$ & 0.359 &  0.13(5)      \\
				& $2_2^- \rightarrow  1_1^-$ & $M1$ & 0.103 &  0.07(3)      \\
				& $2_2^- \rightarrow  2_1^-$ & $M1$ & 0.042 &  0.032(12)      \\
				& $2_2^- \rightarrow  0_1^-$ & $E2$ & 0.165 &  0.13(4)      \\
				& $3_1^- \rightarrow  2_1^-$ & $M1$ & 0.373 &  0.055(16)      \\
				& $4_1^- \rightarrow  2_1^-$ & $E2$ & 1.876 &  1.2(3)      \\
				& $7_1^+ \rightarrow  5_1^+$ & $E2$ & 0.328 &  1.25(8)      \\

				&  &  &   &     \\	 
				$^{207}$Tl 	& $3/2_1^+ \rightarrow  1/2_1^+$ & $M1$ & 0.00 & 0.013(3)       \\
				& $3/2_1^+ \rightarrow  1/2_1^+$ & $E2$ & 2.467 &  2.7(7)       \\
				& $5/2_1^+ \rightarrow  1/2_1^+$ & $E2$ & 2.467 &        \\

				&  &  &   &    \\
				$^{208}$Tl	& $4_1^+ \rightarrow  5_1^+$ & $M1$ & 1.904 &  2.1(3)      \\
				& $5_2^+ \rightarrow  4_1^+$ & $M1$ & 0.146 &  $<$4.7      \\
				& $5_2^+ \rightarrow  5_1^+$ & $M1$ & 0.032 &  $<$1.3      \\
				& $4_2^+ \rightarrow  4_1^+$ & $M1$ & 0.181 &       \\
				& $4_2^+ \rightarrow  5_1^+$ & $M1$ & 0.660 &      \\
				& $3_1^+ \rightarrow  4_1^+$ & $M1$ & 0.018 &      \\

				&  &  &   &    \\
				$^{209}$Tl 	& $3/2_1^+ \rightarrow  1/2_1^+$ & $M1$ & 0.0005 &       \\
				& $5/2_1^+ \rightarrow  3/2_1^+$ & $M1$ & 0.28 &       \\
				& $7/2_1^+ \rightarrow  3/2_1^+$ & $E2$ & 1.0 &        \\
				& $9/2_1^+ \rightarrow  7/2_1^+$ & $M1$ & 1.03 &        \\
				& $9/2_1^+ \rightarrow  5/2_1^+$ & $E2$ & 1.7 &        \\
				& $13/2_1^+ \rightarrow  9/2_1^+$ & $E2$ & 1.3 &   4.2(16)  \\
				& $17/2_1^+ \rightarrow  13/2_1^+$ & $E2$ & 0.347 &   1.2(4)  \\
				
				&  &   & &   \\
				$^{210}$Tl 	& $4_1^+ \rightarrow  5_1^+$ & $M1$ & 2.16  &       \\
				& $5_2^+ \rightarrow  4_1^+$ & $M1$ & 0.039 &       \\
				& $5_2^+ \rightarrow  5_1^+$ & $M1$ & 0.002 &       \\
				& $4_2^+ \rightarrow  4_1^+$ & $M1$ & 0.089 &      \\
				& $4_2^+ \rightarrow  5_1^+$ & $M1$ & 0.188 &      \\
				& $3_1^+ \rightarrow  4_1^+$ & $M1$ & 0.0027 &      \\
				\hline
				\hline

		\end{tabular}}
	\end{center}
\end{table}


{\bf $^{209}$Tl:}  Comparisons of the experimental excitation energies for the positive and negative parity states of $^{209}$Tl with the predictions of shell model calculations are shown in  Fig. \ref{209Tl}. The low-lying states in $^{209}$Tl can be described as two-particle, one-hole configurations outside the doubly magic $^{208}$Pb core. At low excitation energies, all states are occupied by the lowest two valence neutron states above the $N$ = 126 shell gap, i.e., the $g_{9/2}$ orbital with a proton hole in the $s_{1/2}$ orbital, including the $J^\pi$ = ${17/2}^+$ isomer. Except for the states $J^\pi$ = ${3/2}^+$, ${1/2}^+$, which arise from the configuration $\pi d_{3/2}^{-1} \nu g_{9/2}^{2}$. The configuration $\pi s_{1/2}^{-1} \nu g_{9/2}^{2}$ changes to $\pi d_{3/2}^{-1} \nu g_{9/2}^{2}$ for the ${7/2}^{+}$ state in without core excitations calculation. The ${5/2}^+$ state is at higher energy from both sets of calculations with the predominant configuration $\pi d_{5/2}^{-1} \nu g_{9/2}^{2}$. The high-spin states ${21/2}^+$ and ${19/2}^+$ are from the configuration $\pi s_{1/2}^{-1} \nu i_{11/2}^1 g_{9/2}^{1} $. In the calculated state, no core excitations can be seen.

\subsection{Electromagnetic Properties}

In this section, we have discussed the electromagnetic properties for Tl isotopes, reported in Table \ref{t_trans}. For $^{205}$Tl, the calculated $M(1)$ transition values are satisfactory, while the $E(2)$ transition values are less than the experimental data. Although the states included in the transitions are almost at equal energies from all sets of calculations, the discrepancies might be due to the wave-function, which can be tested with the inclusion of the core excitations. For $^{206}$Tl, the calculated $M(1)$ and $E(2)$ transition values are in good agreement with the experimental data. For $^{207}$Tl, all the states above single-particle levels are core-excited states. We have not calculated the electromagnetic transition values with the core-excited state. The calculated $E(2)$ value for the isomeric state ${17/2}^+$ in $^{209}$Tl supports the isomeric nature with hindered transition value. We have also reported electromagnetic properties corresponding to different transitions where experimental data are not available, which might be useful for future experiments.

\section{Conclusions}
\label{summary}
In this work, we have done a comprehensive study of the high-spin states in the $^{204-210}$Tl isotopes with a considerably large model space containing core excitations configurations across the $Z$ = 82, $N$ = 126  shell gap motivated by the recent experimental data. We have reproduced the experimental levels of these isotopes for both the low-lying and high-spin yrast states. We have also compared our calculated electromagnetic transition values with the available experimental data. Our calculations suggest that many of the high-spin states have the main configurations of neutron core excitations across the $Z$ = 82, $N$ = 126 shell gap. In addition, we have predicted some of the core-excited states. It is confirmed that the excitation of neutrons across $N$ = 126 sub-shell closure is essential for the description of the high-spin level structure of Tl isotopes. It is anticipated that these core excitation characteristics will be observed in future experiments and may play a key role in the Pb-region.

\section*{Acknowledgment}
We acknowledge financial support from MHRD, the Government of India, and SERB (India), CRG/2022/005167. We would like to thank the National Supercomputing Mission (NSM) for providing computing resources of ‘PARAM Ganga’ at the Indian Institute of Technology Roorkee, implemented by C-DAC and supported by the Ministry of Electronics and Information Technology (MeitY) and Department of Science and Technology (DST), Government of India.


\end{document}